\documentclass[10pt,letterpaper, aps,twocolumn]{revtex4}
\usepackage{times}
\usepackage{textcomp}
\usepackage{amsmath}
\usepackage{graphicx} 
\usepackage{dcolumn}  
\usepackage{bm}       
\usepackage[english]{babel}

\graphicspath{{}}

\begin{document}

\title[Detecting Neutral Atoms on an Atom Chip]{Detecting Neutral Atoms on an Atom Chip}

\author{Marco Wilzbach%
  \footnote{Corresponding author\quad E-mail:~\textsf{wilzbach@physi.uni-heidelberg.de}}}
\author{Albrecht Haase%
  \footnote{Present address: Institut de Ci\`{e}ncies Fot\`{o}niques ES-08860 Castelldefels (Barcelona), Spain.}}
\author{Michael Schwarz%
  \footnote{Present address: Institut f\"ur Technische Physik, Forschungszentrum Karlsruhe, D-76344 Eggenstein-Leopoldshafen, Germany.}}
\author{Dennis Heine}
\author{Kai Wicker}
\author{S\"onke Groth}
\author{Thomas Fernholz%
  \footnote{Present address: Van der Waals-Zeeman Instituut, Universiteit van Amsterdam, 1018 XE Amsterdam, The Netherlands.}}
\author{Bj\"orn Hessmo}
\author{J\"org Schmiedmayer}
\affiliation{Physikalisches Institut, Universit\"at Heidelberg, D-69120 Heidelberg, Germany.}
\author{Xiyuan Liu}
\author{Karl-Heinz Brenner}
\affiliation{Lehrstuhl f\"ur Optoelektronik, Universit\"at Mannheim, D-68131 Mannheim, Germany.}

\date{Published online August 4, 2006}

\begin{abstract}
Detecting single atoms (qubits) is a key requirement for implementing
quantum information processing on an atom chip. The
detector should ideally be integrated on the chip. Here we present
and compare different methods capable of detecting neutral atoms on
an atom chip. After a short introduction to fluorescence and
absorption detection we discuss cavity enhanced detection of single
atoms. In particular we concentrate on optical fiber based detectors
such as fiber cavities and tapered fiber dipole traps. We discuss
the various constraints in building such detectors in detail along
with the current implementations on atom chips. Results from
experimental tests of fiber integration are also described. In
addition we present a pilot experiment for atom detection using a
concentric cavity to verify the required scaling.
\end{abstract}

\maketitle

\section{Introduction}
Detecting single neutral atoms state selectively is one of the
essential ingredients for developing atomic physics based quantum
technologies, and a prerequisite of most quantum information
experiments. The key question to be answered is how to perform such
measurements using a robust and scalable technology. Ultra cold
atoms can be trapped and manipulated using miniaturized atom chips
\cite{Fol02}. So far, atom detection on atom chips is still based
on unsophisticated methods. Fluorescence detection and absorption
imaging yield detection sensitivities not higher than 10\%. The
detection is also restricted to ensembles of atoms.

On an atom chip micro fabricated wires and electrodes generate
magnetic and electric fields that can be used to trap and
manipulate neutral atoms \cite{Fol02,EJP05}. The atoms are
trapped a few micrometers above the chip surface with high
precision and well defined positions. Many components for
integrated matter wave technology have been demonstrated. Examples
are atom guides \cite{Fol00}, combined magnetic/electrostatic
traps of various geometries \cite{Kru03}, motors and shift
registers \cite{Han01}, atomic beam splitters
\cite{Cas00,Mul00,Sch05,Hom05}, and Bose-Einstein
condensation (BEC) in micro traps \cite{Ott01,Hae01,Sch03}.

Integration of the detection optics on the atom chip is a natural
development for integrated matter wave manipulation. Using
macroscopically sized detectors, state sensitive detection with
near unity efficiency has been realized, see for instance
\cite{Pin00} and references therein. Miniaturizing and integrating
these highly sophisticated and efficient detectors is a
difficult challenge.

The scope of this article is to explore how one can use the
techniques of integrated miniaturized optics for atom detection as
proposed in \cite{Hor03,Lon03,Lev04,Ros04,Eri05} and how to adapt and
implement real world devices to the atom chip and to on-chip atom
detection.
\section{Detecting single atoms}
Single charged particles can be detected through their direct electric
interaction with their environment. Neutral atoms however
interact rather weakly with the environment. The coupling of
electromagnetic radiation to the atom is characterized
by the scattering cross section $\sigma_\mathrm{scat}$
\begin{eqnarray}
    \sigma_\mathrm{scat} &=& \frac{\sigma_\mathrm{abs}}{1+4 \Delta^2},
    \label{eq:AtomLight1}
\end{eqnarray}
where $\sigma_\mathrm{abs}=3\lambda^2/2\pi$ is the resonant
scattering cross section \footnote{For the D2 line of rubidium the on resonant scattering
cross section is given by $\sigma^{\mathrm{Rb}}_\mathrm{abs}=0.29$
\textmu m${}^2$.}
and $\Delta$ the normalized detuning
given by $\Delta=\frac{\omega-\omega_0}{\Gamma}$. $\Gamma$ is the
atomic linewidth.
The real part of the refractive index results in a phase shift
$\phi$ on the light field. For far off-resonant light fields this becomes
\begin{eqnarray}
    \phi &=& \frac{n_\mathrm{at} \sigma_\mathrm{abs}}{4 \Delta},
    \label{eq:AtomLight2}
\end{eqnarray}
where $n_\mathrm{at}$ is the atomic column density.
Atoms can be detected by monitoring the
scattered light in fluorescence, in absorption, or via the imprinted phase shift on the
probing light.
The efficiency of the detection depends on the coupling strength
between the atom and the light, so it is of significant interest to
optimize the atom-photon interaction. This requires a high degree
of control over the optical fields as well as the dynamics of
the atom.

In the following we will restrict ourselves to $n_\mathrm{at}
\sigma_\mathrm{abs}\ll 1$. This is the case for detection
of small atom numbers and
for resonant scattering far below atomic saturation.
\subsection{Measuring the scattered light: Fluorescence Detection}
One way of detecting an atom is to observe its fluorescence. The
basic idea is to drive a transition of the atom to an excited
state with an external light field and to detect the spontaneously
emitted photons. In principle one scattered photon is sufficient
to detect a single atom. The sensitivity of the detection and its
fidelity depend on the detection efficiency of the scattered
light and the suppression of background noise.

Modern photo detectors have a near unity efficiency, and atom
detection is limited by the amount of light collected. Therefore
high numerical aperture collection optics is desirable. In a
realistic setting the collection efficiency $\eta_\mathrm{coll}$
is much less than unity and it is essential to let each atom
scatter many photons. This is best achieved by driving closed
transitions.
Using a model of a two-level atom with
a small number of excitations, the
signal to noise ratio for single atom
detection is given by:
\begin{equation}
    \mathrm{SNR_{fl}}
    = \eta_\mathrm{coll} \sigma_\mathrm{abs} \sqrt{\frac{I_\mathrm{in}\tau}{A \; f_\mathrm{b} + \eta_\mathrm{coll} \sigma_\mathrm{abs}}},
    \label{eq:flDet:SNR}
\end{equation}
where $I_\mathrm{in}$ is the incident photon flux density and $\tau$ the measurement time.
The number of collected fluorescence photons is to be assumed
$n_\mathrm{coll}=\eta_\mathrm{coll} \sigma_\mathrm{abs} j_\mathrm{in}\tau \gg 1$.
$f_\mathrm{b}$ is the fraction of incoming light which is
scattered by anything else in the beam and collected by the
fluorescence detector. $A$ is the size of the detection region
imaged on the detector, with a total background of $A f_\mathrm{b}$ .

In principle, there is no fundamental limit to the efficiency of a
fluorescence detector as long as the atom is not lost from the
observation region. As an example: Using a NA 0.5 (F:1)
optics $\eta_\mathrm{coll}=6.25\%$ and driving the D2 line of Rb
($\sigma^{Rb}_\mathrm{abs}=0.29$ \textmu m$^2$) with about 1/10
saturation intensity ($\sim 10^7$ photons s$^{-1}$ \textmu
m$^{-2}$) one collects about 20 fluorescence photons in 100
\textmu s. Imaging a 10 \textmu m$^2$ detection region onto a
photo detector one needs a suppression factor $f_b$ of $\sim
10^{-3}$ for single atom detection with a $\mathrm{SNR_{fl}} \sim
3$. For trapped ions it is possible to reach very high detection
fidelities. See for example Ref.~\cite{Lei04}. Detection of cold
single neutral atoms has been demonstrated in many experiments.
For instance, up to 20 atoms trapped in a magneto-optical trap
could be counted with a bandwidth of 100 Hz \cite{Mir03}. In
Sec.~\ref{sec:absandflu} we discuss the prospects for atom chip
based fluorescence detectors.

The disadvantage of fluorescence detection is the destructive
nature of the process. The internal state of the detected atom
will be altered, and heating of atoms due to spontaneous emission
is almost unavoidable.

\subsection{Measuring the driving field}

While fluorescence detection uses the spontaneously emitted light,
the presence of an atom will also influence the driving field.
This is described by the susceptibility of the atom. The imaginary
part of the susceptibility describes the absorption, and the real
part the phase shift on the driving field. By measurements on the
driving field a complete atomic signature can be collected.

\subsubsection{Absorption on resonance}
The atomic density can be measured by monitoring the attenuation
of the driving field. In the unsaturated case, this situation is
described by Lambert-Beer law. The transmitted light intensity is
given by:
\begin{eqnarray}
    \frac{I_\mathrm{trans}}{I_\mathrm{in}}=\exp(- n_\mathrm{at} \sigma_\mathrm{abs})
    \sim 1-n_\mathrm{at}\sigma_\mathrm{abs}.
    \label{eq:absorption}
\end{eqnarray}
The absorption signal of Eq.~(\ref{eq:absorption}) provides a direct measure of the
atomic column density. For on resonant excitation on the D2 line
of Rb a column density of one atom per \textmu m${}^2$ leads to
about 30\% absorption. If the mean intensity of the incoming beam
is known, the main uncertainty is determined by measuring the
transmitted light. The signal-to-noise ratio for an absorption
measurement is
\begin{eqnarray}
  \nonumber    \mathrm{SNR_{abs}}&=&\sqrt{j_\mathrm{in}\tau}
    \frac{1-\exp(- n_\mathrm{at}\sigma_\mathrm{abs})}{\sqrt{\exp(- n_\mathrm{at}\sigma_\mathrm{abs})}}\\
    &\approx&\sqrt{j_\mathrm{in}\tau}n_\mathrm{at}\sigma_\mathrm{abs}.
    \sim\sqrt{N_\mathrm{abs} n_{\mathrm{at}} \sigma_{\mathrm{abs}}},
    \label{eq:absorption_SN}
\end{eqnarray}
where $N_\mathrm{abs}=j_\mathrm{in}\tau n_{\mathrm{at}}\sigma_{\mathrm{abs}}$.
The minimum number $n_\mathrm{min}$ of atoms that can be detected with a
$\mathrm{SNR}_\mathrm{abs}=1$ is then given by:
\begin{equation}
    n_\mathrm{min}  = \frac{A}{\sigma_\mathrm{abs}} \frac{1}{\sqrt{j_\mathrm{in}\tau}}
                    = \frac{1}{\sigma_\mathrm{abs}}\sqrt{\frac{\mathrm{A}}{I_\mathrm{in}\tau}},
    \label{eq:absorption_Nmin}
\end{equation}
where we introduced $j_\mathrm{in}$ as the incoming photon flux
($j_\mathrm{in}=A \; I_\mathrm{in}$).
Driving the D2 of Rubidium with an incoming intensity of $10^7$ photons s${}^{-1}$ \textmu m${}^{-2}$ (about
1/10 saturation intensity for Rb atoms) and a waist of 2.5 \textmu m
\mbox{($A=\frac{\pi}{4} w_0^2 \sim 5$ \textmu m${}^2$)} one finds
$n_\mathrm{min} \sim \mathrm{2.5 }\,\tau^{-1/2}$ where $\tau$ is
given in \textmu s.

Equation (\ref{eq:absorption_SN}) assumes that the mean incident photon flux is known to
high precision. In absorption imaging it is common practice to relate the
absorption of the atoms to a reference image by dividing the two
images. Then the noise of the reference image has to be fully
considered.  For small absorption, the signal-to-noise ratio of
Eq.~(\ref{eq:absorption_SN}) is reduced by a factor $1/\sqrt{2}$.
This increases the necessary minimum number of detectable atoms by a factor
$\sqrt{2}$.

To reach unity detection efficiency in absorption imaging it seems
natural to reduce the beam waist as much as possible as indicated
by Eq.~(\ref{eq:absorption_Nmin}). This is however not a
successful strategy as pointed out by van Enk
\cite{vEnk00,vEnk01,vEnk04}. A strongly focused beam is not optimally
overlapped with the radiation pattern of the atoms. The absorption
cross section for such a strongly focused beam becomes smaller
with a decreased spot size.

Another strategy to improve the sensitivity is to increase the
measurement time. The scattered photons will however heat up the
atoms and expel them from the observation window.
If the atom is held by a dipole trap, the measurement time can be
increased significantly (See section \ref{sec:absandflu}). There
is no fundamental limit to the detection of a single atom via
absorption, if it can be kept localized long enough.
\subsubsection{Refraction}
For large detunings ($\Delta\gg 1$) the
absorption decreases as $\propto
\Delta^{-2}$ as can be seen from Eq.~(\ref{eq:AtomLight1}). In addition the transmitted beam acquires a phase
shift $\phi_\mathrm{atom}$. This is caused by the refractive index of the
atoms. The phase shift decreases only with $\phi_\mathrm{atom} \propto
\Delta^{-1}$ as shown by Eq.~(\ref{eq:AtomLight2}). The phase-shift has been used to image clouds of
cold atoms and include Mach-Zehnder type interferometers
\cite{Kad01}, for phase-contrast imaging \cite{And96}, or in line
holography by defocussed imaging \cite{Tur04}.

The minimal detectable phase shift $\Delta\phi$ in an
interferometer is given by the number / phase uncertainty relation
$\Delta\phi\Delta N = 1$ resulting in $\Delta\phi_\mathrm{min}=
1/\sqrt{j_\mathrm{in}\tau}$ (neglecting absorption).
From these scaling laws for the scattered light and the phase
shift one finds for the signal-to-noise ratio for
dispersive atom detection:
\begin{eqnarray}
    &&\mathrm{SNR_{disp}}
    =
    \frac{\phi}{\Delta\phi}
 \sim\sqrt{N_\mathrm{scat} n_{\mathrm{at}} \sigma_{\mathrm{abs}}},
    \label{eq:phase_SN}
\end{eqnarray}
with $N_\mathrm{scat}=j_\mathrm{in}\tau n_{\mathrm{at}}\sigma_{\mathrm{scat}}$.
$\mathrm{SNR_{disp}}$ depends only on the total number of scattered photons
$N_\mathrm{scat}$. In fact it
is the same as for on resonant detection, and the optimal
achievable SNR for classical light as shown by Hope \emph{et al.}
\cite{Hop04,Lye04,Hop05}. Going off resonance does not help in
obtaining a better measurement compared to plain absorption.

The off-resonant detection of atoms however has significant
advantages in combination with cavities to enhance the interaction
between photons and the atom \cite{Hop04}. Non-destructive and
shot noise limited detection becomes possible.
\subsection{Cavities}
Cavities enhance the interaction between the light and atoms.
The photons are given multiple chances to interact with
an atom located in the cavity.
The number of interactions is enhanced
by increasing the number of round trips.
The latter is determined by the cavity
finesse $n_{\mathrm{rt}}=\mathcal{F}/2\pi$. This can be used to
improve the signal-to-noise ratios. Even with moderate
finesse cavities single atom detection can be achieved.

\subsubsection{Absorption on resonance}

The probability for absorption during each round trip is
determined by the ratio between the atomic absorption cross
section $\sigma_{\mathrm{abs}}=3\lambda^2/2\pi$ and the beam
cross section inside the cavity $A=\frac{\pi}{4}w_0^2$. A natural
figure of merit for cavity assisted absorption is therefore
\begin{equation}
    C_1=\frac{\mathcal{F}}{2\pi}\frac{\sigma_{\mathrm{abs}}}{A}=\frac{3\lambda^2}{\pi^3}\frac{\mathcal{F}}{w_0^2}.
    \label{eq:Purcell}
\end{equation}
This quantity is identical to the cooperativity parameter
$C_1=g_0^2/2\kappa\Gamma$ \cite{Ber94}, which relates the single
photon Rabi frequency of the atom-photon system $g_0$ to the
incoherent decay rates of the cavity field $\kappa$ and atomic
excitation $\Gamma$. Interestingly, a reduced
cavity mode waist can compensate for a small cavity finesse
\cite{Hor03}.

When the cooperativity parameter is smaller than unity and the
atomic saturation is low, the signal-to-noise ratio for shot-noise
limited single atom detection becomes
\begin{equation}
    \mathrm{SNR} = \sqrt{j_\mathrm{in}\tau}\frac{\kappa_T}{\kappa}C_1
                 = \frac{3\lambda^2}{\pi^3}\sqrt{j_\mathrm{in}\tau}\frac{\kappa_T}{\kappa}\frac{\mathcal{F}}{w_0^2}
                 \sim \sqrt{\frac{\kappa_T}{\kappa} N_\mathrm{abs} C_1},
    \label{eq:Purcell2}
\end{equation}
where $N_\mathrm{abs}=j_\mathrm{in}\tau \frac{\sigma_{\mathrm{abs}}}{A} \frac{\kappa_T}{\kappa} n_{\mathrm{rt}}$,
$\kappa_T$ is the
mirror transmission rate, and $\kappa$ the total cavity decay rate
\cite{Hor03}. For a fixed measurement time an increased
signal-to-noise ratio can be obtained by increasing the
cooperativity parameter. This can be done by increasing the cavity
finesse, or by decreasing the beam waist.
\subsubsection{Refraction}
Similarly, the phase shift induced by the atom in the cavity
increases with the each round trip.  Accordingly the signal to
noise ratio Eq.~(\ref{eq:phase_SN}) is then increased to
\begin{equation}
    \mathrm{SNR}= \sqrt{j_\mathrm{in}\tau}\frac{\kappa_T}{\kappa}\frac{C_1}{\Delta} \sim \sqrt{\frac{\kappa_T}{\kappa} N_\mathrm{scat} C_1}
    \sim \sqrt{\frac{\kappa_T}{\kappa} N_\mathrm{abs} C_1}/\Delta,
    \label{eq:Purcell3}
\end{equation}
where
$N_\mathrm{scat}=j_\mathrm{in}\tau \frac{\sigma_{\mathrm{scat}}}{A} \frac{\kappa_T}{\kappa} n_{\mathrm{rt}}$.
When the cooperativity parameter is larger than unity, non-destructive detection
with low photon scattering becomes possible \cite{Hop04}.
\subsubsection{Many atoms in a cavity}

The above considerations concern the coupling of a single atom to
a cavity. This can be extended to the many-atom case by
introducing a many-atom cooperativity parameter
$C=N_\mathrm{eff}C_1$, where $N_\mathrm{eff}$ is an effective
number of atoms in the cavity mode, which takes into account the
spatial dependence of the coupling constant $g(\vec{r})=g_0
\psi(\vec{r})$, given by the cavity mode function $\psi(\vec{r})$,
and the atomic density distribution $\rho(\vec{r})$. The fraction $N_\mathrm{eff}$
of the total atom number $N$, which are maximally coupled to the
cavity mode, is given by the overlap integral of both functions
\begin{equation}
    N_\mathrm{eff}=N \int d^3r\,\rho(\vec{r})|\psi(\vec{r})|^2.
    \label{eq:Neff}
\end{equation}
The absorptive and dispersive effects of the atoms on the cavity
amplitude \cite{Hor03} scale linearly with this effective atom
number as long as the atomic saturation is low.
\subsection{Concentric cavity}
As seen above, a small mode diameter $w_0$ is advantageous.  Such a
small $w_0$ can be obtained by using a near-concentric cavity.
Consider the case when the cavity is formed by two identical
mirrors with radius of curvature $R$ separated by a distance $L$.
The beam waist $w_0$ in the cavity center is given by
\begin{equation}
    w_{0}^2=\frac{\lambda}{2\pi}\sqrt{L(2R-L)}.
    \label{eq:waist}
\end{equation}
The concentric geometry occurs when the mirror separation $L$
approaches the value $2R$. The waist size $w_0$ becomes small but
the beam size on the cavity mirrors
\begin{equation}
    w^2=\frac{R\lambda}{\pi}\sqrt{\frac{L}{2R-L}},
    \label{eq:spot}
\end{equation}
becomes large as one approaches the concentric limit. A large
mirror spot size requires very uniform mirrors, as deviations from
a spherical mirror shape will lower the optical finesse
drastically. Furthermore, as the concentric point is approached,
the cavity also becomes extremely sensitive to misalignments and
vibrations. For more details on cavities see Siegman
\cite{Sie86}.

\subsection{Miniaturization}
The principle advantage of miniaturizing the cavities is that for
a fixed geometry, i.e. for a constant ratio of R and L, the mode
diameter scales with size (see Eqns.~(\ref{eq:waist}) and
(\ref{eq:spot})).  This automatically increases the cooperativity
parameter.  In fact $C_1$ scales like $C_1 \propto
w_0^{-2}$ and high values for $C_1$ can be reached even
for moderate finesse as illustrated in Table.~\ref{tab:finesses}.
\begin{table}
\begin{tabular}{|c|c|c|c|c|c|c|}\hline
$L$         & $w_0$      & $\mathcal{F}$ & $g_0$             &$g_0/\Gamma$
                                                                    &$g_0/\kappa$
                                                                           & $C_1$\\\hline
[\textmu m] & [\textmu m]&               & $2\pi\times$[MHz] &  & & \\
\hline
200         & 30         & 20000          & 12               & 4.0 & 0.65  & 1.3\\
50          & 7.5        & 5000           & 97               & 32  & 0.32  & 5.2\\
20          & 5          & 5000           & 230              & 77 & 0.31 & 11.8\\
10          & 2.5        & 5000           & 650              & 217   & 0.43  & 47\\
10          & 2.5        & 1250           & 650              & 217   & 0.11  & 11.8\\
10          & 2.5        & 250            & 650              & 217   & 0.02  & 2.3\\
$10^4$      & 2.5        & 250            & 21                & 6.9  & 0.68  & 2.3\\
$10^5$      & 2.5        & 250            & 6.5               & 2.2 & 2.2  & 2.3\\
$10^5$      & 2.5        & 50             & 6.5               & 2.2 & 0.43  & 0.47\\
\hline
\end{tabular}
  \centering
  \caption{Properties of various cavities as a function of mode geometry and finesse.
  The cooperativity parameter $C_1$ stays high if the mode waist is kept small even at
  moderate cavity finesse. For small mode volumes the single photon Rabi frequency can
  become very high.}
  \label{tab:finesses}
\end{table}
Miniaturization allows to build cavities with a very small mode
volume $V_m=\int d^3\vec{r}|\psi(\vec{r})|^2$. A small cavity
volume has the advantage that photons will interact more strongly
with atoms localized inside the cavity as the light intensity per
photon increases. The interaction between the photon and the atom
is described by the atom-photon coupling constant (single photon
Rabi frequency) $g_0=\sqrt{3\Gamma c \lambda^2/(4\pi V_m)}$. It
determines how much the dressed atomic energy levels inside the
cavity are shifted by the presence of a single photon. As $g_0$ is
increased, single atom - single photon coupling becomes feasible if
$g_0\gg(\kappa,\Gamma)$ can be satisfied.

Since for a fixed finesse the cavity line width $\kappa$ becomes
larger with decreasing length $L$ ($\kappa \propto L^{-1}$) the
fulfillment of the above condition for strong coupling and CQED
does not improve as dramatically as the cooperativity parameter
$C_1\sim w_0^{-2}$. One has $g_0/\Gamma\sim 1/w_0\sqrt{L}$ and
$g_0/\kappa\sim\sqrt{L}/w_0$ so the length must be chosen precisely
to achieve strong coupling (see Table ~\ref{tab:finesses}). It is however
always an advantage to choose $w_0$ small.

For detection of atoms the increased cooperativity parameter is the main benefit from miniaturization.
Miniaturized cavities with moderate finesse can detect single atoms. Therefore miniaturization is much
more beneficial for atom detection than for cavity QED experiments.
\section{Properties of Fiber Cavities}
The advantage of fibers is that they can be easily handled using well established techniques.
For instance, transfer mirror coatings can be directly glued to the fibers.
Different fibers can be melted together using commercially available fusion splicers.
The fabrication of fiber optical components does usually not require expensive and time consuming lithographic techniques.
Fibers also have the advantage that they have very low optical loss, usually less than 3 dB/km.

\begin{figure}[h!]
    \includegraphics[width=0.45\textwidth]{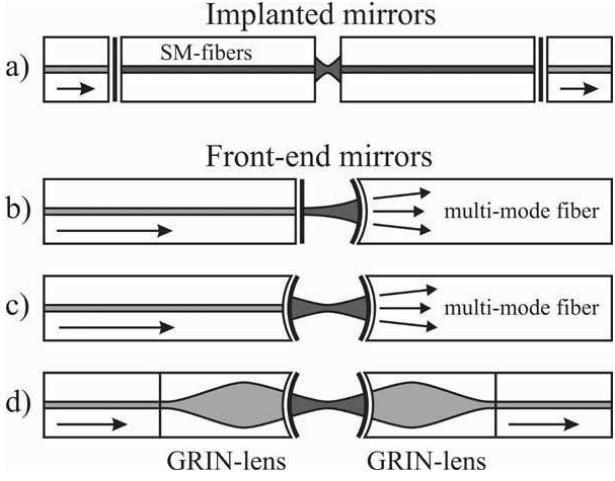}%
    \caption{a) A cavity with implanted mirrors. This cavity requires little alignment of the gap, but the
    mode volume is large. This configuration can be realized using only single mode fibers.
    b) A plano-concave cavity. Incoupling from a single mode fiber is efficient, but the
    mode diameter becomes large at the location of the atom. Outcoupling to a single mode fiber is not optimal, but a multi-mode fiber
    can collect the transmitted light well.
    c) A concave-concave cavity. The mode diameter can be controlled,
    but coupling to single mode fibers is not optimal. d) Focussing optics can be integrated to optimize the fiber-cavity coupling
    efficiency. Cavities b) - d) require active alignment of the gap.}
    \label{fig:variouscavities}
\end{figure}

For miniaturization of optical components for atomic physics several proposals on atom chip integration have been
presented \cite{Hor03,Lev04,Ros04,Mok04,Arm03,Liu05}.
In our approach optical fibers are attached to the chip to form fiber-based cavities and to make small dipole traps.
For fiber cavity experiments it is interesting to put atoms into the optical resonator.
To place the atom inside the cavity it is possible to cut the fiber into two pieces and place the atom in the fiber gap.
It is also possible to use a hollow fiber and guide the atom inside it \cite{Key00}.
Because it is quite difficult to load atoms into a hollow fiber, especially
if it is mounted on a micro-structured surface, the following section deals with fiber-gap-fiber configurations as
the ones outlined in Fig.~\ref{fig:variouscavities}.

One realization of a fiber cavity is to implant mirrors into the fibers, such as Bragg gratings or
dielectric coatings inserted between fiber pieces. See Fig.~{\ref{fig:variouscavities}a).
This has the advantage that the interaction volume (the gap where the atoms pass) is defined only by the bare fiber tips.
Thus only the fiber itself has to be mounted on the atom chip with little or no need for alignment actuators.
Mirrors and tuning actuators can be placed far from the gap.
This allows a higher integration of the optics.
A disadvantage of this method is that the cavity becomes quite long and that the extra interfaces increase the optical loss of the cavity.
Using a long cavity will increase the mode volume and therefore also reduce $g_0$.
For atom detection the length of the cavity is fortunately not a very important parameter, as indicated by Eq.~(\ref{eq:Purcell2}).
\subsection{Loss mechanisms for a cavity}
The quality of an optical resonator is described by its finesse.
The finesse is related to the losses by:
\begin{equation}
    \mathcal{F} = \frac{\pi}{\sum\limits_{i} \alpha_i},
    \label{eq:Finesse}
\end{equation}
where $\alpha_i$ are the loss probabilities for the different loss channels.
For example, a cavity mirror transmittance of $T=0.01$ corresponds to $\alpha=0.01$.
As long as $\Sigma_i\alpha_i\ll 1$ the cavity finesse is accurately calculated by Eq.~(\ref{eq:Finesse}). For further details on
resonators we again refer to \cite{Sie86}.
For atom detection it is desirable to obtain a high signal-to-noise ratio.
This does not automatically mean that the finesse should be maximized.
The signal-to-noise ratio for atom detection in the weak coupling and low atomic saturation is given by Eq.~(\ref{eq:Purcell2}),
where $\kappa_T\propto\alpha_T$ describes the mirror transmittance and $\alpha_\mathrm{tot}=\alpha_T+\alpha_\mathrm{other}$
is the total cavity loss. This implies that $\mathrm{SNR}\propto\alpha_T/(\alpha_T+\alpha_\mathrm{other})^2$ which is maximized for
$\alpha_T=\alpha_\mathrm{other}$. For each kind of cavity there is an optimal choice for the mirror transmittance.

When the cavity contains a gap (See Fig.~\ref{fig:variouscavities}a) the main loss mechanism is related to the reduced coupling efficiency across the gap.
This loss depends mainly on three things.
(1) It is clear that the mode matching of the two fibers is not optimal and losses
will occur as the gap is traversed. This loss increases with the gap size.
(2) Also, in the case of transverse misalignment of the two fibers modes are not well-matched at
the fiber facets.
(3) A tilt between the two fibers also increases the mismatch between the two modes.
In this case the loss also increases with the length of the fiber gap.
\begin{figure}[h!]
    \includegraphics[width=0.4\textwidth]{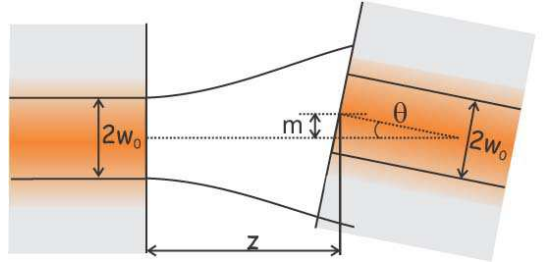}%
    \caption{The fiber light mode diverges,
    after leaving the fiber on the left with a mode field diameter of $2 w_0$. The second fiber on the right has
    a longitudinal displacement of $z$, a transversal displacement of $m$ and an angular misalignment of $\theta$. }
    \label{fig:gapmisalignment}
\end{figure}

Figure \ref{fig:gapmisalignment} sketches these misalignments of the gap.
The power coupling efficiency $\eta$ for two fibers can be calculated in the paraxial approximation
by the overlap integral of the involved transverse mode functions.
\begin{equation}
    \eta=\frac{\left|\int \!\!\! \int_{\mathbf{A}} \Psi_{1}(\vec{r}) \Psi_{2}^{*}(\vec{r})dA \right|^2}{
    \int\!\!\!\int_{\mathbf{A}} \Psi_{1}(\vec{r}) \Psi_{1}^{*}(\vec{r})dA
    \int\!\!\!\int_{\mathbf{A}} \Psi_{2}(\vec{r}) \Psi_{2}^{*}(\vec{r})dA},
    \label{eq:lossdef}
\end{equation}
where $\Psi_1(\vec{r}),\, \Psi_2(\vec{r})$ are the fiber optical field modes, and $\mathbf{A}$ is the
plane coinciding with one of the fiber facets.
In this text we use this measure to describe cavity losses.

\subsection{Losses due to the gap length} It is obvious that the losses increase with expanding gap size $z$.
The light exiting the first fiber at $z=0$ diverges until it hits the opposite fiber at a distance $z$.
We can describe the light leaving the first fiber as
\begin{eqnarray}
\nonumber    \Psi_1(x,y,z)=\sqrt{\frac{2}{\pi}}\frac{1}{w(z)}\exp{\left(-\frac{ik\left( x^2+y^2 \right)}{2 R(z)}\right)} \times\\
   \exp{\left(-\frac{x^2+y^2}{w(z)^2}\right)}\exp{(ikz-i\Phi(z))},
    \label{eq:gaussmode1}
\end{eqnarray}
where $w_0$ is the Gaussian waist at the first fiber facet,
$R(z)=z+z_0^2/z$ the radius of curvature, $w(z)=w_0 \sqrt{1+z^2/z_0^2}$ the beam radius,
$z_0=\pi w_0^2/\lambda$ the Rayleigh length, and $\Phi(z)=\arctan \left(z/z_0\right)$ the Gouy-phase.
The wavelength is $\lambda=\lambda_0/ n$ and the wavenumber $k=k_0 n$
in the medium between the fibers with refractive index $n$. The wavelength and the wavenumber in vacuum are $\lambda_0$ and $k_0$, respectively.
Assuming that the two fibers are identical, the mode function at the facet of the second fiber is
\begin{equation}
    \Psi_2=\Psi_1(x,y,z=0)=\sqrt{\frac{2}{\pi}}\frac{1}{w_0}\exp{\left(-\frac{x^2+y^2}{w_0^2}\right)}.
    \label{eq:gaussmode2}
\end{equation}

The loss $\alpha_{\mathrm{gap}}$ due to the length of the gap can be calculated from the overlap
between functions~(\ref{eq:gaussmode1}) and (\ref{eq:gaussmode2}).
\begin{eqnarray}
\nonumber    \alpha_{\mathrm{gap}}(z) &=& 1- \left| \int\limits_{-\infty}^{\infty}\int\limits_{-\infty}^{\infty}\Psi_1(x,y,z) \Psi_2^*(x,y,0) dx\, dy \right|^2\\
    &=&\frac{z^2}{4 z_0^2+z^2}\approx\left(\frac{z}{2z_0}\right)^2=\left(\frac{\lambda^2}{4\pi^2w_0^4}\right)z^2.
    \label{eq:gaploss}
\end{eqnarray}
Here the transverse and angular misalignments are assumed to be zero.
A plot of a finesse measurement with varying gap size $z$ is shown in Fig.~\ref{fig:gapscan-su8}a.
In this case mirrors with a loss of 1\% are assumed to be located inside the fibers.
This loss is decreased if the gap between the fibers is small and the fiber has a small numerical aperture i.e. a large $w_0$.
However, to detect atoms $w_0$ should be small, as indicated by Eq.~(\ref{eq:Purcell2}).

\subsection{Losses due to transversal misalignment}
If the two fibers are transversally misaligned to each other there will also be a mode mismatch when the light is coupled between the fibers.
This loss can easily be calculated from Eq.~(\ref{eq:lossdef}).
Using normalized Gaussian mode functions for two identical fibers (with $z=0$) with waist $w_0$ choosing the gap and angular
misalignment to be zero one has
\begin{eqnarray}
\nonumber    \Psi_1 (x,y,0)&=& \sqrt{\frac{2}{\pi}} \frac{1}{w_0}\exp\left( -\frac{x^2+y^2}{w_0^2} \right),\\
\nonumber    \Psi_2 (x,y,0)&=& \Psi_1 (x-m,y,0)\\
    &=&\sqrt{\frac{2}{\pi}} \frac{1}{w_0}\exp\left( -\frac{(x-m)^2+y^2}{w_0^2} \right),
\end{eqnarray}
where $m$ is the transverse misalignment of one of the fibers. This gives a loss
due to the transversal misalignment $\alpha_{\mathrm{tra}}$:
\begin{eqnarray}
\nonumber    \alpha_{\mathrm{tra}}(m) &=& 1- \left| \int\limits_{-\infty}^{\infty} \int\limits_{-\infty}^{\infty} \Psi_1(x,y,0) \Psi_2^*(x,y,0) dx\, dy \right|^2\\
    &=&1- \exp \left(-\frac{m^2}{w_0^2} \right) \approx \frac{m^2}{w_0^2},
    \label{eq:transloss}
\end{eqnarray}
where the last approximation is valid only for $m \ll w_0$. A plot of a finesse measurement with varying
transversal misalignment is shown in Fig.~\ref{fig:gapscan-su8}b.

\begin{figure}[h!]
    \includegraphics[width=0.4\textwidth]{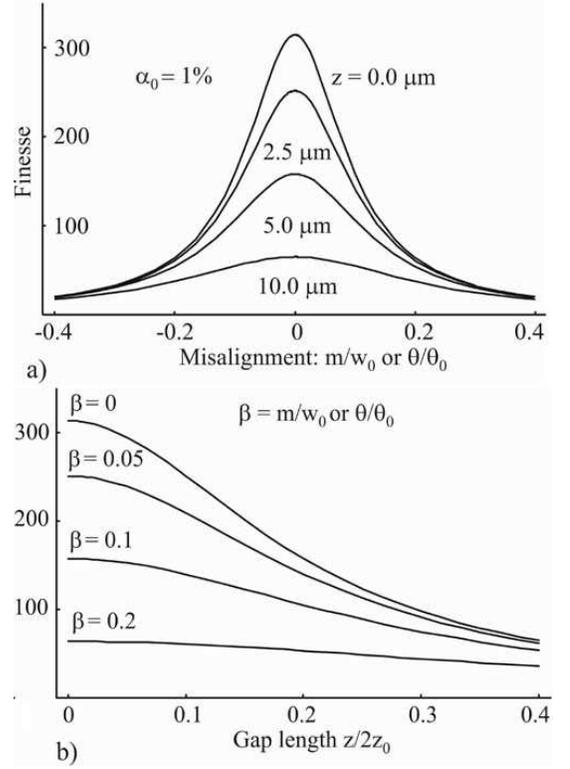}
    \caption{Theoretical calculations were conducted for an implanted mirror cavity.
    The fiber mode diameter was $w_0=2.5$ \textmu m at a wavelength of 780 nm.
    All losses except the gap losses were set to 1\%.
    Plot a) shows the finesse as a function of $\theta$ and $m$, which
    were normalized to $\theta_0$ and $w_0$, respectively.
    $\theta_0$ is given by $\lambda /(\pi w_0)=w_0/z_0$.
     The different curves correspond to different gap lengths starting with the highest
    finesse at $z=0$ \textmu m and ending with the lowest at $z=10$ \textmu m.
    In b) the finesse is plotted as a function of $z/2z_0$.
    The different curves correspond to different $\beta$, where $\beta$ is either $\theta /\theta_0$ or $m /w_0$. $\beta$ ranges from $0$ to $0.2$}
\label{fig:theorygapfinesse}
\end{figure}

\subsection{Losses due to angular misalignment}
Another kind of loss emerges from angular misalignment of the two optical axes
of the fibers. This is calculated by performing a basis
change of Eq.~(\ref{eq:gaussmode1}) and evaluating the overlap given by Eq.~(\ref{eq:lossdef}).
The mode leaving the first fiber (see
Fig.~\ref{fig:gapmisalignment}) is approximately described by
\begin{equation}
    \Psi_1(x,y,z) \approx \sqrt{\frac{2}{\pi}}\frac{1}{w_0} \exp{\left(
    -\frac{x^2+y^2}{w_0^2}  \right)} \exp{(i k z)},
\end{equation}
for small $z\ll z_0$ neglecting diffraction effects. For a small rotation $\theta$ around the
y-axis is given by
\begin{eqnarray}
\nonumber
    \left(
    \begin{array}{c}
        x\\
        y\\
        z
    \end{array}
    \right)
    &=&
    \left(
    \begin{array}{ccc}
        \cos\theta  & 0 & \sin\theta \\
        0           & 1 & 0 \\
        -\sin\theta & 0 & \cos\theta
    \end{array}
    \right)
    \left(
    \begin{array}{c}
        x'\\
        y'\\
        z'
    \end{array}
    \right)\\
    &\approx&
    \left(
    \begin{array}{ccc}
        1       & 0 & \theta \\
        0       & 1 & 0 \\
        -\theta & 0 & 1
    \end{array}
    \right)
    \left(
    \begin{array}{c}
        x'\\
        y'\\
        z'
    \end{array}
    \right).
\end{eqnarray}
Transforming $\Psi_1(x,y,z)$ into the coordinate system $(x', y', z')$
the mode function for the incident beam at the
input plane $z'=0$ of the second fiber becomes
\begin{equation}
    \Psi_1(x',y',z'=0) \approx \sqrt{\frac{2}{\pi}}\frac{1}{w_0} \exp{\left(
    -\frac{x'^2+y'^2}{w_0^2}  \right)}\exp{(i k x' \theta)}.
    \label{eq:angpsi1}
\end{equation}
The mode function for the second fiber at the input plane $z'=0$
is given by
\begin{equation}
    \Psi_2(x',y',z'=0) = \sqrt{\frac{2}{\pi}}\frac{1}{w_0} \exp{\left(
    -\frac{x'^2+y'^2}{w_0^2}  \right)}.
    \label{eq:angpsi2}
\end{equation}
The loss $\alpha_{\mathrm{ang}}$ due to a pure angle misalignment can be
calculated from the overlap between functions (\ref{eq:angpsi1}) and (\ref{eq:angpsi2}).
\begin{eqnarray}
\nonumber
    \alpha_{\mathrm{ang}}(\theta) &=& 1- \left| \int\limits_{-\infty}^{\infty}\int\limits_{-\infty}^{\infty}\Psi_1(x',y', 0) \Psi_2^*(x',y', 0) dx'\, dy' \right|^2\\
    &=&1-\exp \left( -\frac{\pi^2 w_0^2}{\lambda^2} \theta^2 \right)\approx
    \left( \frac{\theta}{\theta_0}\right)^2,
    \label{eq:angloss}
\end{eqnarray}
with $\theta_0=\lambda/(\pi w_0 )= w_0/z_0$.
In general these misalignments are not independent of each other. The combined formula for all the misalignment losses
is given by \cite{Sar79}:
\begin{eqnarray}
\nonumber
    \alpha(\theta,m,z)=1-\eta\approx 1-\mu(z)\times\\
    \exp \left[ -\mu(z) \left(
    \frac{m^2}{w_0^2}+\frac{\pi^2 \theta^2 w_0^2}{\lambda^2}+\frac{\theta^2 z^2}{2 w_0^2}-\frac{m\theta z}{w_0^2}
    \right) \right],
    \label{eq:totloss}
\end{eqnarray}
with $\mu(z)=4z_0^2/(4z_0^2+z^2)$.
A setup with transversal and longitudinal displacement and angle misalignment is plotted in Fig.~\ref{fig:gapscan-su8}.
One experimental advantage is that the loss rate depends quadratically on small misalignments.
\subsection{Fresnel reflections}
Another effect that has to be taken into account is the reflection of light at the interfaces between the fibers and the introduced gap.
It results from the step in the refractive index and is called Fresnel reflection.
For a typical fiber $(n_{1}=1.5)$ in air or vacuum $(n_{2}=1)$ a fraction of $R_{F}=(n_{2}-n_{1})^{2}/(n_{2}+n_{1})^{2}=4\% $ of
the incident light is reflected from a polished flat connector.
In the case of a fiber gap cavity, these additional reflections lead to a system of three coupled resonators.
While for any system of coupled resonators stationary solutions exist, i.e. eigenmodes of the electromagnetic field, they will in general not be
equidistantly distributed in the frequency domain.
\begin{figure}[h!]
    \includegraphics[width=0.45\textwidth]{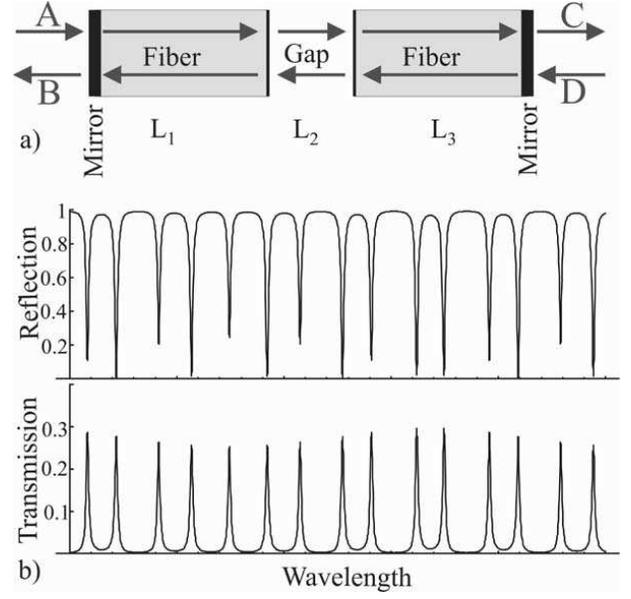}%
    \caption{a) Schematic description of the cavity formed by the three coupled resonators.
    b) Reflectance and transmittance of a cavity formed by three coupled resonators.
    The system is constituted by \mbox{Mirror-Fiber-Gap-Fiber-Mirror} and has $\alpha=0.05$.
    In addition $L_1\ne L_3$ and $L_2=\lambda/4$ plus an integer multiple of $\lambda/2$. The period of the
    envelope is given by the length difference of the two fiber pieces. }
    \label{fig:fresnel}
\end{figure}

In a simple one-dimensional model, the coupled cavity system can be treated as a chain
of transfer matrices \cite{Sie86}.
Figure ~\ref{fig:fresnel} shows the calculated
reflected and transmitted intensities for a model resonator of
rather low finesse.

The light intensity inside the gap depends on the specifics of the mode.
This changes the overall losses, as can be seen in Fig.~\ref{fig:fresnel}b.
Consequently the reflected and transmitted intensities vary from mode to mode. The effect is strongest
when the gap length is a multiple of $\lambda / 2$.  In experiments coupling atoms to the light field it
has to be considered that the intensity inside the gap depends on the order of the mode and on the exact
length of the gap. In absorption experiments only the relative change of transmitted light intensity is
measured, which remains mode independent for light intensities well below saturation.
Never the less it leads to a mode dependent signal-to noise ratio. Using fibers with $n$=1.5 and intensities
below atomic saturation the SNR changes by roughly 15\%
\section{Other fiber optical components for the atom chip}
Not only cavities can be mounted on the surface of an atom chip.
Miniaturization offers various advantages for a couple of other optical components.
Lenses with extremely short focal lengths can be used to focus light to small areas.
Such devices can be used to generate very tight dipole traps \cite{Sch01,Sch02,Rey03}, or to collect light from very well-defined volumes \cite{Eri05}.
\subsection{Fluorescence and absorption detectors\label{sec:absandflu}}
A fluorescence detector can be built using a single mode tapered
lensed fiber and a multi-mode fiber. In this case, the tapered
fiber is used to optically pump the atom. The multi mode fiber is
used to collect the fluorescent light. For this configuration one
can put the multi mode fiber at any angle with respect to the
tapered lensed fiber. At small angles, atoms located in the focus
of the tapered lensed fiber will scatter light \emph{away} from
the multi mode fiber. This would constitute an absorption
detector. Such a device is illustrated in
Fig.~\ref{fig:chipdetectors}c). If the angle between the two fibers
is sufficiently large only the fluorescent emission from the atom
will be \emph{collected} by the multi mode fiber. A picture of a
$90^{\circ}$ fluorescence detector setup is shown in
Fig.~\ref{fig:chipdetectors}b). If no atom is present in the beam
focus, very little light is scattered into the fiber. As soon as
an atom is present, this signal increases. For single atom
detection these two techniques are not sensitive enough, because photon recoils will expel the atom from the detection region
and the multi mode fiber only collects a few photons. Therefore one must also trap the
atom(s) in the tapered lensed fiber focus. This can be
accomplished by a dipole trap.
\subsection{A single mode tapered lensed fiber dipole trap}
To increase the number of detectable photons, the atom can be
trapped inside the area where it interacts with the light. One
possibility is to hold the atom in place with a dipole trap.
An atom with transition frequency $\omega_0$ is attracted to the intensity maximum of a red-detuned laser beam ($\omega<\omega_0$).
A focussed laser beam forms an atom trap in the beam focus.
For a far-detuned dipole trap quantities such as
trap depth, lifetime and number of scattered photons can be estimated easily.
A detailed derivation of dipole trap parameters discussed in the following paragraph can be found in \cite{Gri00}.
The potential depth of a dipole trap is
\begin{equation}
    U_{\mathrm{dip}}(\vec{r})=\frac{\pi c^2\Gamma}{2 \omega_0^3}\left( \frac{2}{\Delta_{D2}}+\frac{1}{\Delta_{D1}}\right)
     I(\vec{r}),
\end{equation}
and the scattering rate is given by
\begin{equation}
    \Gamma_{\mathrm{sc}}(\vec{r})=\frac{\pi c^2\Gamma^2}{2 \hbar \omega_0^3}
    \frac{}{}\left( \frac{2}{\Delta_{D2}^2}+\frac{1}{\Delta_{D1}^2}\right)I(\vec{r}),
\end{equation}
where $\Delta_{D2}$ and $\Delta_{D1}$ are the detunings of the laser with respect to the D2 and D1 lines of the alkali atom.
The formulas apply to alkali atoms with a laser detuning much larger than the hyperfine structure splitting and a
scattering rate far from saturation. It follows that the scattering rate for a given potential can be decreased by
choosing a larger detuning and a higher laser intensity.
The potential depth scales as $\sim\Gamma/\Delta$, whereas the scattering rate scales as $\sim\Gamma^2/\Delta^2$.

The D2 line of Rubidium, \mbox{${}^2S_{1/2}\rightarrow {}^2P_{3/2}$},
has a wavelength of 780 nm, and the D1 line
\mbox{${}^2S_{1/2}\rightarrow {}^2P_{1/2}$} has a wavelength of 795 nm. A
standard tapered lensed fiber generates a typical waist of
$w_0=2.5$ \textmu m at a wavelength of 780 nm. Far off resonance
dipole traps can be realized with easy to use, high power diode
lasers at a wavelength of 808 nm. Standard single mode lasers diodes are
available with a power up to 150 mW.
Assuming a coupling efficiency into the fiber of around 20\%, one gets 30 mW of laser power in the dipole trap.
These parameters yield a trap depth of 3.9 mK, and a transverse (longitudinal) trap frequency of 80 kHz  (6 kHz).
The heating rate for the dipole light is almost negligible.
Experiments have shown that the lifetime is mainly limited by the background pressure of the vacuum chamber \cite{Mil93}.

Using resonant light to detect an atom increases the number of scattered photons as well as
the heating rate enormously. In this case the trap lifetime drops
to below a millisecond, which yields several thousands scattered
photons before the atom is lost.

A multi mode fiber as detector at a distance of 40 \textmu m can collect
2\% of the fluorescent emission (See Fig.~\ref{fig:chipdetectors}b). Assuming a detector efficiency of 60\% (APD-based photon
counter), including additional losses in the rest of the optical
detector setup, it should be possible to detect about 100 photons
during the trap lifetime. The signal-to-noise ratio is difficult to calculate without knowing the background noise.

For an absorption detector, such as the one illustrated in Fig.~\ref{fig:chipdetectors}c, it is possible to use Eq.~(\ref{eq:absorption_SN}) to calculate
the expected signal-to-noise ratio. In this case one finds SNR=4 for $\tau=100$\textmu s and $I_\mathrm{in}=\mathrm{10}^7$ photons $s^{-1}$\textmu m${}^{-2}$.
\section{Integration of fibers on the atom chip\label{sec:SU8}}
A method for mounting fibers accurately is needed to integrate the above devices with atom chips.
Fibers can be held at the right place with the help of some kind of fiber grippers, aligned and finally glued to the chip.
This strategy will not work well for fiber cavities because they need to be actively aligned during operation (See Fig.~\ref{fig:variouscavities}b-d).
For active alignment of an on-chip cavity an actuator is needed.
Useful actuators, such as piezoelectric stacks, are quite large compared to the fibers. Typical dimensions exceed 2x2x2mm.
This places restrictions on the level of optical integration one can obtain on the chip.
Components that don't need active alignment, such as the cavities with implanted mirrors (See Fig.~\ref{fig:variouscavities}a)
or tapered fibers for dipole traps, can be directly integrated on the chip.
To obtain a high degree of precision in the mounting we have developed a lithographic method where a thick photoresist is patterned to form the
mounting structures \cite{Liu05}.
\subsection{Building fiber cavities}
We have so far realized cavities with curved and planar mirrors at the front-end
of two fibers, and cavities with implanted mirrors made out of one piece of fiber with
a gap for the atoms similar to the ones shown in Fig.~\ref{fig:variouscavities}.

The mirror coatings are attached to the fibers using a \emph{transfer technique}.
In this process, dielectric mirror coatings with a transmittance of 0.1\% to 1\% are manufactured on a glass substrate.
The substrate is either flat or contains an array of spherical microlenses for curved coatings.
The adhesion between the coating and the glass substrate is fairly low.
When the coating is glued to an optical fiber it is possible to transfer the
coating from the substrate to the fiber \cite{Jena}. A series of pictures showing this transfer is found in Fig.~\ref{fig:transfertechnique}.

The transmittance T is chosen to be around 1\% for cavities with implanted mirrors (Fig.~\ref{fig:variouscavities}a) because of the higher loss.
For the front-end mirrors (Fig.~\ref{fig:variouscavities}b-d), the loss is lower therefore it is more suitable to choose a
mirror with transmittance around 0.1\% to obtain a high SNR for the detection of atoms.

The higher loss for cavities with implanted mirrors has several origins in addition to the ones caused by the gap.
For some of our cavities the main intrinsic loss sources are the glue layers holding the mirror coatings to the fibers.
These glue layers are typically a few micrometers thick.
The transversal confinement of the fiber is absent in the glue layers as well as in the mirror coatings.
This leads to losses similar to the ones discussed for the gap length, i.e Eq.~(\ref{eq:gaploss}). Glued mirrors are
also sensitive to angular misalignment, which can be treated by Eq.~(\ref{eq:angloss}).
\begin{figure}[h!]
    \includegraphics[width=0.48\textwidth]{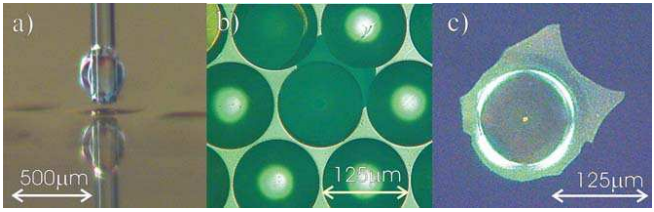}
    \caption{
    Picture a) shows a fiber with a drop of glue on the tip. The fiber is
    hovering 100 microns above a micro lens with dielectric mirror coating. The lower part is a mirror image of the real fiber.
    Once the fiber is glued to the coating it can be removed by gently pulling the fiber away from the substrate.
    In picture b) one can see the micro lens array from above.
    A closer look reveals that coatings are missing from two of the lenses.
    One of these is attached to the fiber shown in c).
    Additionally in c) light was coupled in from the other side of the fiber, so one can see the fiber
    core in the center of the picture.}
    \label{fig:transfertechnique}
\end{figure}

Curved coatings may be used to build a concave cavity with front-end mirrors.
With this design we reached a finesse of greater than 1000 as shown in Fig.~\ref{fig:innercavity}.
Complications due to Fresnel reflections do not exist for the front-end cavity.
The quality of the mirrors and their alignment determine how high the finesse may become.
The mode profile of the optical fiber does usually not match the cavity mode. This leads to a poor incoupling into the
front-end cavity. This problem can be solved by transferring the mirror coatings to a tapered fiber, or
a small grin lens to focus the light into the cavity mode.
This unfortunately leads to a more complicated and error prone fabrication process.
In addition, a cavity with front-end mirrors must be mounted on alignment actuators for efficient tuning of the cavity.
This makes miniaturization and integration of multiple cavities harder.
\begin{figure}[h!]
    \includegraphics[width=0.4\textwidth]{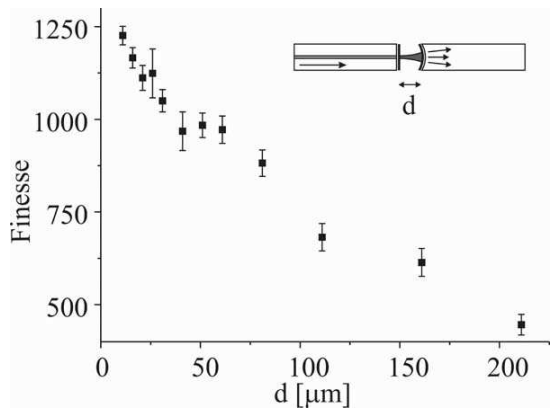}%
    \caption{Finesse measurement of a cavity with a flat mirror on a single mode
    fiber on the incoupling side and a curved mirror with $r=350$ \textmu m on a multimode fiber on
    the outcoupling side, as the one illustrated in Fig.~\ref{fig:variouscavities}b. Both mirrors have an reflectance of R=99.9\%.
    The finesse stays above 400 for fiber separations smaller than 200 \textmu m and
    has a maximal value of 1200 for small mirror separations.}
\label{fig:innercavity}
\end{figure}

A cavity with the geometry shown in Fig.~\ref{fig:variouscavities}a) was built by inserting two planar mirrors into a fiber with length about
10 cm. This fiber cavity is subsequently cut into two halves and the new surfaces are polished.
This cavity has the advantage that it is very easy to align and mount.
An actual cavity of this kind is shown in Fig.~\ref{fig:chipdetectors}d) where the cavity is mounted on an atom chip using a SU-8 structure
to hold the fibers.
The drawback with this cavity geometry is that the finesse is rather low compared to the front-end mirrors, mainly because of the fiber
gap and the glue layers as described above. Cavities with inserted mirrors typically reach a finesse of a few
hundred instead of a few thousand as for the front-end cavities. The losses due to the glue layers could be reduced by direct coating of the
fiber instead of using the transfer technique.
The losses due to the gap itself could also be reduced by introducing collimation optics in
the gap, such as a small grin lens or a tapered fiber. Such additional optics will however introduce additional Fresnel reflections and may also
require active alignment of the gap, as for the front-end cavity.
\begin{figure}[h!]
    \includegraphics[width=0.4\textwidth]{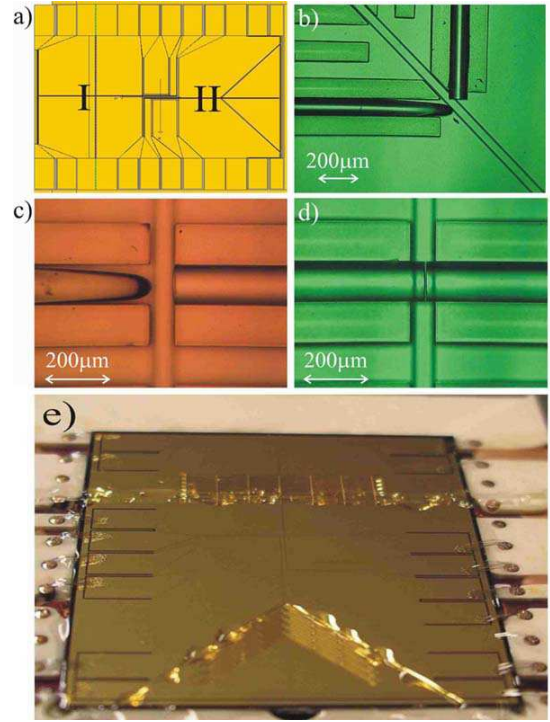}
    \caption{A drawing of the actual atom chip is shown in a). The pictures b)-d) come from areas
    I and II.
    In b) the image shows
    the $90^{\circ}$ fluorescence fiber detector setup. It is built by a tapered fiber (the horizontal fiber)
    used for a dipole trap and optical pumping. The vertical fiber is a multimode fiber used to collect scattered
    light.
    Picture c) shows a tapered fiber facing a multimode fiber. This setup can
    be used for absorption detection of a few atoms.
    Image d) shows a fiber cavity with a gap of a few microns. The actual atom chip containing the structures b)-d) is shown
    in e)
    }
    \label{fig:chipdetectors}
\end{figure}
\subsection{The SU-8 resist}
To mount fibers on the atom chip a lithographically patterned photoresist called SU-8 was used.
SU-8 is an epoxy based negative resist with high mechanical, chemical and thermal stability.
Its specific properties facilitate the production of thick structures with very smooth, nearly vertical sidewalls \cite{Liu05}.
It has been used to fabricate various micro-components. Examples include optical planar waveguides with high thermal stability
and controllable optical properties.
Mechanical parts such as microgears \cite{Sei02} for engineering applications, and microfluidic systems for chemistry \cite{Duf98} have also been built.
The SU-8 is typically patterned with 365-436 nm UV-light.

To assess the quality of the alignment structures, we use the SU-8 structures to hold fiber optical resonators.
The finesse of the resonator strongly depends on losses introduced by misalignment as described by Eq.~(\ref{eq:totloss}).
The layout of the used alignment structure with fibers is shown in Fig.~\ref{fig:su8structures}.
This design includes funnels to simplify assembly.
To avoid angular misalignment the total length of the alignment structure is quite long (6000 \textmu m).
The structure is divided into several subsegments to reduce stress induced by thermal expansion.

\begin{figure}[h!]
    \includegraphics[width=0.48\textwidth]{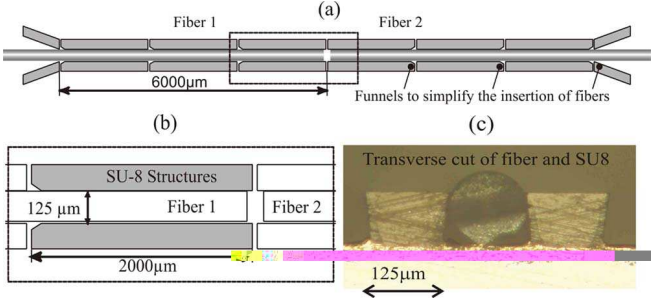}
    \caption{a) Layout of the alignment structure and a
    magnified part (dotted rectangle) in b). c) Fiber in a SU-8 structure
    mounted on a gold coated atom chip. The atom chip and the SU-8 structures have
    been cut with a wafer saw. The SU-8 maintains structural integrity during the cutting.}
    \label{fig:su8structures}
\end{figure}
\subsection{Test of the SU-8 structure}
The quality of the SU-8 fiber splice was determined by measuring the finesse of a mounted resonator.
The finesses of two intact fiber resonators were found to be $\mathcal{F}_{1}=110.4 \pm 0.3$ and $\mathcal{F}_{2}=152.8 \pm
1.1$. After cutting the fiber and polishing the new surfaces, the two fiber pieces were introduced into the SU-8 structures.
We observed the fiber ends under a microscope and minimized the gap size.
For these resonators the finesses were found to be $\mathcal{F}_{1}=101.1 \pm 0.5$ and $\mathcal{F}_{2}=132.0 \pm
1.3$. This corresponds to an additional average loss of $\alpha=(0.29 \pm 0.04$)\%.
Neglecting other additional losses, this corresponds to a transversal misalignment of $m=150$ nm or an
angular misalignment of $\theta=6.3\times 10^{-3}$ rad $\approx 0.36^{\circ}$.
To test thermal stability the temperature of the substrate was varied between 20 \textcelsius\ and 70
\textcelsius. The finesse of the inserted fiber resonator showed no change during heating.
\begin{figure}[h!]
    \includegraphics[width=0.35\textwidth]{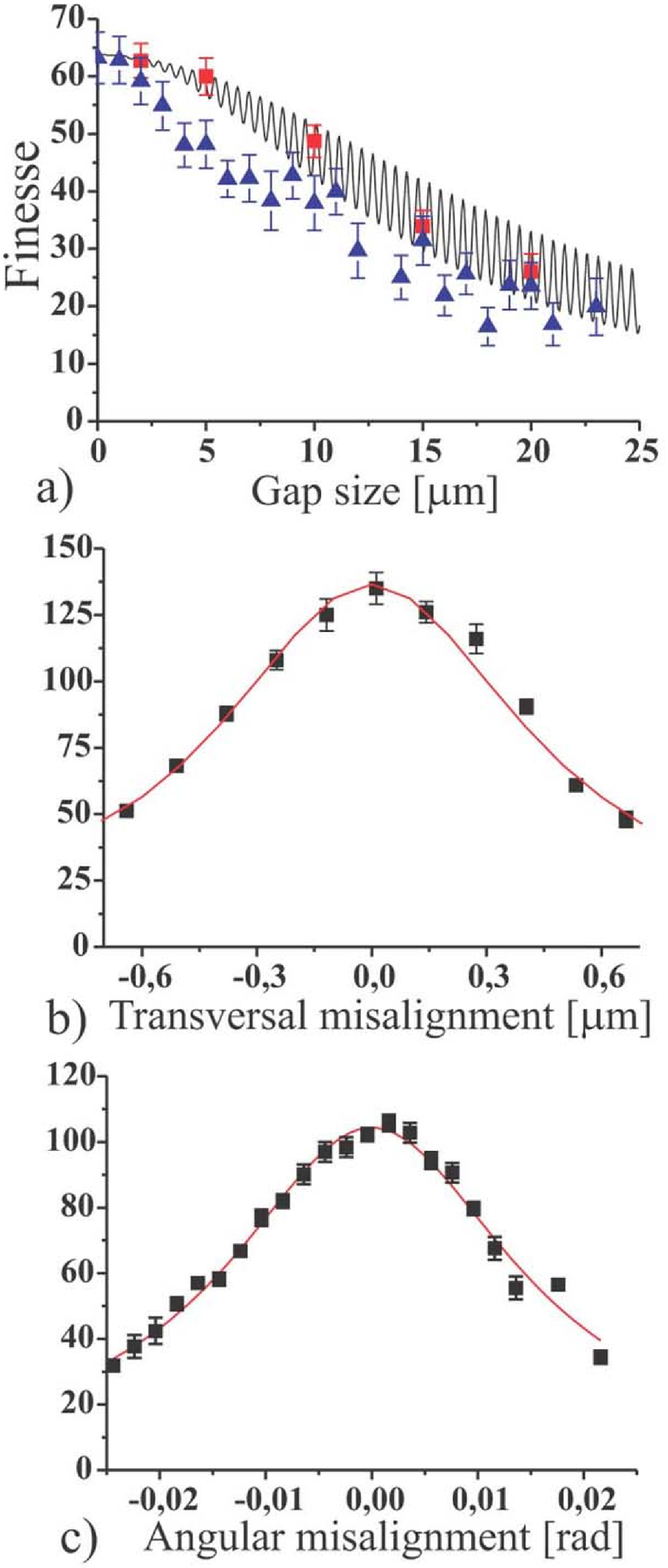}
    \caption{
    Measurement of the finesse of a fiber cavity as a function of the gap size is shown in a).
    The oscillating function shows the theoretical curve for the finesse with varying gap size. The oscillation of the
    finesse is due to the multiple reflections between the facets and the mirrors.
    The red squares correspond to the finesse measured for a cavity mounted inside a SU-8 structure.
    The blue triangles correspond to a finesse measurement where the cavity is held by precision translation stages.
    Graph b) shows the finesse of a different cavity measured as a function of transversal misalignment.
    In c) the finesse is plotted as a function of angular misalignment.}
\label{fig:gapscan-su8}
\end{figure}
Another test for the quality of the SU-8 structures was a finesse measurement as
a function of gap size. One measurement was done in the SU-8 structures and one
outside the structures.
The measurement outside was performed with nanopositioning stages, which in principle can be tuned to a few nanometers.
With the positioning stages the finesse was optimized.
Nevertheless, the finesse obtained inside the SU-8 structures always remained higher than the ones obtained using the positioning stages.
The results of these measurements are shown in Fig.~\ref{fig:gapscan-su8}a).
The structures have also allowed long-term stability of the cavities in a high-vacuum environment.
\section{Pilot test for atom detection with small waists}
To assess the consequences of Eq.~(\ref{eq:Purcell2}) a test using a macroscopic cavity was performed \cite{Haa05b}.
The goal was to detect atoms with a cavity with moderate finesse and a very small mode diameter.
\subsection{Dropping atoms through a concentric cavity}
To explore atom detection using cavities with small waists, we built
a magneto-optical trap (MOT) for
${}^{85}$Rb atoms approximately 20 mm above the center of a near
concentric cavity
with a finesse of about 1100 formed by normal high-reflectance mirrors.
The MOT contained $\sim 10^{7}$ atoms at a temperature of 35 \textmu K.
In a first experiment, we switched off the trap completely and
monitored the atomic cloud as it fell freely through the cavity.
This configuration was used to estimate the detection sensitivity of
the cavity.

\begin{figure}[h!]
    \includegraphics[width=0.36\textwidth]{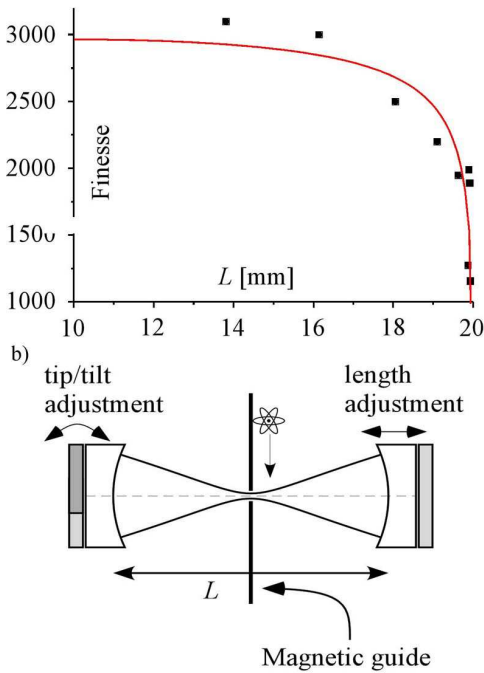}
    \centering
    \caption{a) The finesse of the cavity decreases as the concentric
    point is approached. The curve is calculated from the cavity geometry
    and mirror specifications. b) Schematic description of the cavity
    and the magnetic guide mounted next to the cavity.}
    \label{fig:cavity}
\end{figure}
We measured the light intensity transmitted through the cavity with an
amplified photodiode for high light intensities or with a
photomultiplier tube (PMT) for low light intensities. The PMT
provided a near shot-noise limited detection. The low-noise
electronic amplification limited the detection bandwidth to 20 kHz.
The main source of technical noise in our setup was due to
mechanical vibrations of the vacuum chamber that held the cavity.

To minimize the influence of vibrations and long term drifts one of
the cavity mirrors was mounted on a piezoelectric tripod that allowed
us to keep the TEM${}_{00}$ mode centered on the cavity axis. The second
mirror was mounted on a translating piezoelectric stage for wavelength
stabilization.
Figure \ref{fig:cavity} illustrates how the cavity finesse was
reduced as the concentric point was approached for our cavity formed
by two mirrors with a radius of curvature of $R=10$ mm and a
transmittance $T=10^{-3}$.
For a mirror separation far less than the concentric limit the cavity had
a finesse of 3000. The finesse dropped to 1100 when the
separation was 70 \textmu m from the concentric point. The cavity mode
waist was 12.1 \textmu m for this separation.

The drop in the cavity transmission signal from freely falling
atoms is plotted in Fig.~\ref{fig:freefall}a). The different curves show
different pump powers, corresponding to empty cavity
transmissions between 1 pW and 600 pW. The atom number in the MOT is
$1.5\times10^7$, the signal drops by 90$\%$ as long as the atomic
transition is not saturated (Fig.~\ref{fig:freefall}b). Combining the results of \cite{Hor03}
with Eq.~\ref{eq:Neff} the effective atom number becomes $N_\mathrm{eff}=2.5\pm0.5$. This was
consistent with an independent atom number measurement based on
fluorescence imaging. To determine the sensitivity limit of the
cavity detector, the number of atoms in the MOT was gradually
reduced until the MOT contained $3.5\times10^5$ atoms. This atom number
produced a signal drop of approximately $10\%$. We consider this to be the
sensitivity limit. A theoretical fit results in an effective atom
number sensitivity of $N_\mathrm{eff}=0.1\pm0.05$.
\begin{figure}[h!]
    \includegraphics[width=0.35\textwidth]{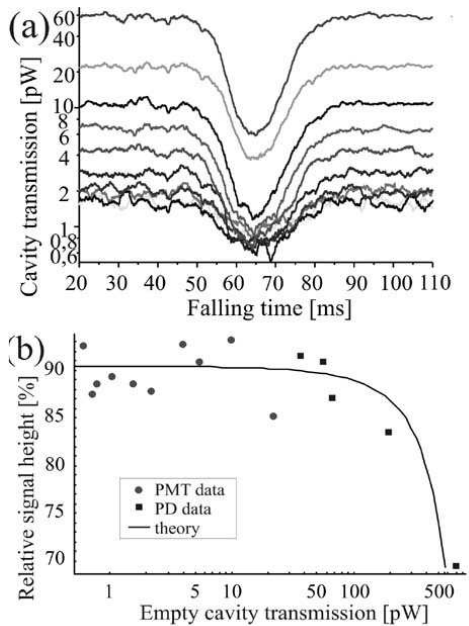}
    \centering
    \caption{(a) Cavity transmission signal for atoms
    dropped from a MOT. Different curves correspond to different cavity pump powers.
    The signal has been averaged over 2.5ms for better visualization.
    (b) Relative drop of the signal due to the atoms in (a). The circles (squares)
    come from measurements with a PMT (photodiode) for different light intensities.
    The black line is calculated numerically.}
    \label{fig:freefall}
\end{figure}
\subsection{Detecting Magnetically Guided Atoms}
As a next step, atoms were magnetically guided to the cavity
center using a wire guide \cite{Den99}. By
changing the current in the guiding wire the overlap between the
atoms and the cavity mode could be adjusted. In
Fig.~\ref{fig:guide} we plot the cavity transmission as the
position of the magnetic guide is varied across the cavity mode.
As the atomic overlap with the cavity mode was increased, we observed an increased drop in cavity transmission.
From the duration of the drop in transmission the temperature of the guided atoms could be determined to be 25 \textmu K.
\begin{figure}[h!]
    \includegraphics[width=0.38\textwidth]{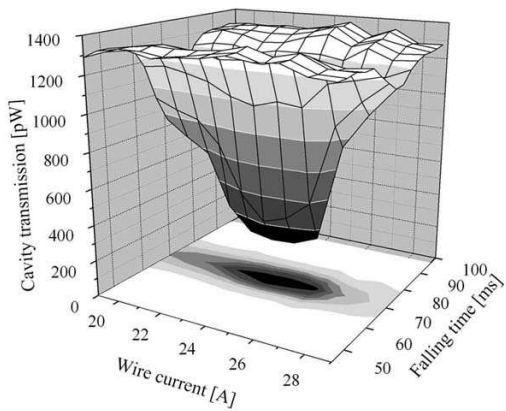}
    \centering
    \caption{Cavity transmission signal from atoms being magnetically
    guided through the cavity mode. The position of the potential minimum is
    linearly dependent on the wire current.}
\label{fig:guide}
\end{figure}
The density distribution for the atoms was much larger than the Rayleigh volume of the cavity.
Consequently, it was not possible to distinguish individual atoms in the guide using our cavity.
This cavity would show a detectable change in the transmission signal if a single atom crossed the region of
maximum coupling as $N_\mathrm{eff}$ can be as small as $0.1$.
The precision in the positioning can be improved using magnetic microtraps produced by atom chip surface traps \cite{Fol02}.
To achieve the same detection sensitivity with a cavity mode diameter of 2 \textmu m a finesse of 40 would be sufficient.
\section{Conclusion}
In this article we have explored and compared various methods of
detecting single atoms on an atom chip. Miniaturization brings
advantages for such detectors. To detect atoms with a cavity, the
requirements on the cavity finesse relaxes quadratically as the
mode diameter of the cavity is decreased. This favorable scaling
law was tested in a pilot experiment where atoms were detected
when magnetically guided through a near-concentric cavity

The simplest way to get to a fully integrated atom detector on an
atom chip is to mount optical fibers on to the surface of the
chip. Such fiber devices can be used for fluorescence, absorption,
and cavity assisted detection of single atoms.
To integrate fibers, we have developed a precise lithographic
method that allows robust positioning of fibers to within <150
nanometers. The high accuracy allows very
reliable fiber-to-fiber coupling, suitable for integration of
fiber cavities (Fig.~\ref{fig:variouscavities}a).
We have built and integrated a fiber
cavity on an atom chip which should be able to detect a single
atom with a SNR of better the 5 within a measurement time of 10
\textmu s. The same lithographic method has been used to position
tapered lensed fibers on the chip. These can be used to combine
dipole traps and fluorescence/absorption detectors.

We acknowledge financial support from the Landesstiftung
Baden-W\"urttemberg, Forschungsprogramm
Quanteninformationsverarbeitung and the European Union, contract
numbers IST-2001-38863 (ACQP), MRTN-CT-2003-505032 (Atom Chips),
and HPRI-CT-1999-00114 (LSF).  The atom chip shown in this work
was fabricated at the The Braun Submicron Center at
Weizmann Institute of Science.
%

\end{document}